\documentclass[aps,prd,reprint,nofootinbib,superscriptaddress,longbibliography]{revtex4-2}

\usepackage{amsmath,amssymb}
\usepackage{mathrsfs}
\usepackage{graphicx}
\usepackage{xcolor}
\usepackage[colorlinks=true,linkcolor=blue,citecolor=blue,urlcolor=blue]{hyperref}
\usepackage{orcidlink}

\begin{document}

\title{Gauss–Bonnet Coupling in the Bondi–Sachs Hierarchy and the Tower of Higher-Spin Charges
}

\author{Maickol Muñoz-Palma\orcidlink{0009-0008-1017-4406}}
\email{m.munoz58@ufromail.cl}
\affiliation{Departamento de Ciencias Físicas, Universidad de La Frontera, Casilla 54-D, 4811186 Temuco, Chile.}
\author{Alvaro Restuccia}
\email{alvaro.restuccia@uantof.cl}
\affiliation{Departamento de F\'isica, Facultad de Ciencias Básicas, Universidad de Antofagasta, Casilla 170, Antofagasta, Chile.}
\author{Francisco Tello-Ortiz \orcidlink{0000-0002-7104-5746}}
\email{francisco.tello@ufrontera.cl}
\affiliation{Departamento de Ciencias Físicas, Universidad de La Frontera, Casilla 54-D, 4811186 Temuco, Chile.}

\date{}

\begin{abstract}
We determine where the coupling of scalar--Gauss--Bonnet gravity enters the Bondi--Sachs description of radiating spacetimes, for an arbitrary coupling function and a radiating scalar field. The effective Gauss--Bonnet stress tensor decays faster than any source able to modify the leading asymptotic structure: the Bondi mass-loss formula and the angular-momentum balance law are exactly those of the minimally coupled Einstein--scalar system, and the smooth expansion at null infinity persists to all orders. The coupling first appears in the balance law of the spin-2 aspect carried by the leading coefficient of the Weyl scalar $\Psi_0$, as a flux built from the scalar charge aspect and the time derivative of the gravitational news. We derive this law without symmetry assumptions and show that the Gauss--Bonnet flux is invariant under supertranslations, although $\Psi_0$ itself is not. When the first derivatives of the coupling function vanish at the asymptotic value of the scalar, the imprint moves up the tower of higher-spin charges by a definite number of steps; for the couplings behind spontaneous scalarization it is linear in the coupling and controlled by nonperturbative scalar charges, while on the branch continuously connected to general relativity it is quadratic. For an equal-mass black-hole binary the flux accumulates secularly during the inspiral, and we give its leading-order value in closed form.
\end{abstract}

\maketitle

\section{Introduction}
\label{sec:intro}

The asymptotic structure of radiating spacetimes at future null infinity $\mathscr{I}^+$, formulated by Bondi, van der Burg, Metzner and Sachs~\cite{Bondi:1962px,Sachs:1962wk}, underlies the modern understanding of gravitational radiation: the Bondi mass-loss formula, the Bondi--Metzner--Sachs (BMS) group, and the flux-balance laws that relate changes of asymptotic charges to radiative fluxes. These balance laws encode the gravitational memory effect~\cite{Zeldovich:1974gvh,Christodoulou:1991cr,Strominger:2014pwa,Pasterski:2015tva} \cite{Nichols:2017rqr} and are used as nonperturbative consistency conditions on gravitational waveforms~\cite{Mitman:2020pbt,Mitman:2020bjf}, a role that becomes central as a first detection of the memory comes within reach~\cite{Zosso:2026detection}.

With the prospect of testing gravity with high-precision waveforms, the same structures have been studied beyond general relativity (GR). The asymptotic solution space, symmetries and memory effects are known for Brans--Dicke and general scalar--tensor theories~\cite{Hou:2020tnd,Tahura:2020vsa}, for Einstein--Maxwell--dilaton theory~\cite{Lu:2019jus,Tan:2020hfa}, and for dynamical Chern--Simons gravity~\cite{Hou:2021oxe,Hou:2021bxz}. For theories with higher-derivative couplings, Heisenberg, Yunes and Zosso showed that the memory formula depends only on the second-order action, so that a scalar--Gauss--Bonnet (sGB) coupling does not modify it~\cite{Heisenberg:2023prj,Heisenberg:2024cjk} ; the same framework has since been applied to Einstein--{\AE}ther and generalized Proca gravity~\cite{Heisenberg:2025aether,Heisenberg:2025proca}. Maibach and Zosso derived the Wald--Zoupas~\cite{Wald:1999wa} supertranslation flux and balance law for luminal Horndeski theories, showed that the Gauss--Bonnet term does not survive the pullback to $\mathscr{I}^+$, and stated that the leading Bondi solution is unaffected by it~\cite{Maibach:2026jbf}. Maibach subsequently integrated the Bondi--Sachs hierarchy for a generic effective stress tensor and derived the falloff bounds that a smooth $\mathscr{I}^+$ requires, applying them to scalar--vector--tensor theories with coupling functionals that are regular at the vacuum; logarithmic functionals were explicitly left for future work~\cite{Maibach:2026afb}. Numerical-relativity waveforms in sGB gravity confirm that the memory is modified mainly indirectly, through the change in the merger dynamics~\cite{Gasparotto:2026}, and the scalar radiation carries a memory of its own~\cite{Zosso:2026scalar}.

Scalar--Gauss--Bonnet gravity, which arises in the low-energy limit of string theory~\cite{Zwiebach:1985uq,Gross:1986mw,Kanti:1995vq} and is the prototype of theories with hairy~\cite{Sotiriou:2013qea,Sotiriou:2014pfa} \cite{Yunes:2011we} or spontaneously scalarized~\cite{Doneva:2017bvd,Silva:2017uqg} \cite{Antoniou:2017acq,Doneva:2022ewd} black holes, sits in a peculiar place with respect to these general results. Its Horndeski representation involves the functionals $G_5\propto f'(\Phi)\ln X$ and $G_4\supset f''(\Phi)X\ln X$~\cite{Kobayashi:2011nu}, which are precisely of the logarithmic type excluded from the classification of Ref.~\cite{Maibach:2026afb}. The questions of where exactly the coupling enters the Bondi--Sachs solution space, with which coefficient, and whether it is compatible with a smooth (log-free) expansion, have therefore not been answered by direct integration. The closest precedent is the analysis by L\"u and Mao of the $D\to4$ limit of Einstein--Gauss--Bonnet gravity, a theory without scalar news, where the mass-loss formula was found to be unaffected and the Gauss--Bonnet term was found to act one order after the integration functions~\cite{Lu:2020mjp}.

In this paper we answer these questions and identify the asymptotic charge through which the coupling first becomes visible. Working directly with the covariant form of the field equations, we compute the asymptotic expansion of the Gauss--Bonnet tensor on a Bondi--Sachs background that carries both tensor and scalar radiation, and follow it through the hierarchy. The effective Gauss--Bonnet stress tensor lies between one and three powers of $1/r$ below the falloffs that could alter the leading asymptotic structure, and its leading coefficients, which we obtain in closed covariant form and without symmetry assumptions, depend only on the shear, the mass aspect and the scalar charge aspect. As a consequence, the mass-loss formula and the evolution of the angular-momentum aspect are exactly those of the minimally coupled Einstein--scalar system, nonperturbatively in the coupling, and the smooth expansion at null infinity persists to all orders for an arbitrary coupling function, because the Gauss--Bonnet source never reaches the orders at which the radial integrations of the hierarchy are resonant.

The coupling becomes dynamical one level higher, in the evolution of the $O(r^{-3})$ coefficient of the angular metric, through the product of the scalar charge aspect and the time derivative of the news; from there it propagates to the hypersurface functions and, at the evolution of the fifth coefficient of the scalar, to the scalar hierarchy. In Newman--Penrose language the new term is a source in the evolution of the leading coefficient of $\Psi_0$, and integrated between nonradiative epochs it becomes a flux in the balance law of the spin-2 charge aspect. Although $\Psi_0$ depends on the BMS frame, this Gauss--Bonnet flux is invariant under supertranslations, so it isolates the contribution of the coupling in a frame-independent way. The same mechanism organizes the whole family of coupling functions: when the first derivatives of the coupling vanish at the asymptotic value of the scalar, the imprint moves up the tower of higher-spin charges by a definite number of steps. For the couplings responsible for spontaneous scalarization it then becomes linear in the coupling and controlled by nonperturbative scalar charges, whereas on the branch connected to GR it is quadratic. Finally, for an equal-mass black-hole binary the flux accumulates secularly during the inspiral, with the angular pattern of the nonlinear memory, and we give its leading-order amplitude in closed form.

We use signature $(-,+,+,+)$ and units $G=c=1$.

\section{Theory, field equations and structure of the Gauss--Bonnet tensor}
\label{sec:setup}

\subsection{Field equations}

We consider the action~\cite{Maibach:2026jbf}
\begin{equation}
S=\frac{1}{2\kappa_0}\int d^4x\,\sqrt{-g}\,\Big[R-\tfrac12(\nabla\Phi)^2+\lambda\,f(\Phi)\,\mathcal{G}\Big],
\label{eq:action}
\end{equation}
where $\mathcal{G}=R_{abcd}R^{abcd}-4R_{ab}R^{ab}+R^2$, $f(\Phi)$ is an arbitrary smooth function and $\lambda$ a coupling with dimensions of length squared. The scalar equation is
\begin{equation}
\Box\Phi=-\lambda\,f'(\Phi)\,\mathcal{G} .
\label{eq:scalar}
\end{equation}
The metric equation can be written in terms of the double-dual Riemann tensor contracted with the Hessian of $f$,
\begin{equation}
G_{ab}=\tfrac12\nabla_a\Phi\nabla_b\Phi-\tfrac14g_{ab}(\nabla\Phi)^2-4\lambda\,H_{ab},
\label{eq:metric}
\end{equation}
with $H_{ab}\equiv P_{acbd}\,\nabla^c\nabla^df(\Phi)$ and
\begin{align}
P_{abcd}={}&R_{abcd}-g_{ac}R_{bd}+g_{ad}R_{bc}-g_{bd}R_{ac}+g_{bc}R_{ad}\notag\\
&+\tfrac12R\,(g_{ac}g_{bd}-g_{ad}g_{bc}).
\label{eq:P}
\end{align}
The tensor $H_{ab}$ obeys the identities
\begin{equation}
\nabla^aH_{ab}=-\tfrac18\,\mathcal{G}\,\nabla_bf,\qquad g^{ab}H_{ab}=-G_{ab}\nabla^a\nabla^bf,
\label{eq:identities}
\end{equation}
the first of which guarantees that the divergence of Eq.~\eqref{eq:metric} vanishes on Eq.~\eqref{eq:scalar}, and fixes the coefficient $-4$. In trace-reversed form,
\begin{equation}
R_{ab}=\tfrac12\nabla_a\Phi\nabla_b\Phi-4\lambda\big(H_{ab}-\tfrac12g_{ab}H^c{}_c\big).
\label{eq:tracerev}
\end{equation}
Writing the equations as $G_{ab}=8\pi T_{ab}$, as in Ref.~\cite{Maibach:2026afb}, the effective Gauss--Bonnet stress tensor is $8\pi\,T^{\rm GB}_{ab}=-4\lambda\,H_{ab}$. Since Eqs.~\eqref{eq:scalar} and \eqref{eq:metric} are linear in $\lambda$, every equation of the asymptotic hierarchy is the sum of its Einstein--scalar part and a term proportional to $\lambda$.

\subsection{Algebraic structure and order counting}
\label{sec:structure}

The double dual is the Weyl tensor plus Ricci terms,
\begin{equation}
P_{abcd}=C_{abcd}-\tfrac12\big(g_{ac}\hat S_{bd}-g_{ad}\hat S_{bc}-g_{bc}\hat S_{ad}+g_{bd}\hat S_{ac}\big),
\end{equation}
with $\hat S_{ab}\equiv R_{ab}-\tfrac13R\,g_{ab}$, so that, with $f_{ab}\equiv\nabla_a\nabla_bf=f'(\Phi)\nabla_a\nabla_b\Phi+f''(\Phi)\nabla_a\Phi\nabla_b\Phi$,
\begin{equation}
H_{ab}=C_{acbd}f^{cd}-\tfrac12\Big(g_{ab}\hat S_{cd}f^{cd}+\hat S_{ab}\Box f-\hat S_{ac}f^c{}_b-\hat S_{bc}f^c{}_a\Big).
\label{eq:Hdecomp}
\end{equation}
On shell the Ricci part is itself determined by the scalar, $R_{ab}=\tfrac12\nabla_a\Phi\nabla_b\Phi+O(\lambda)$.

Consider a null tetrad adapted to Bondi coordinates $(u,r,x^A)$: $l$ tangent to the outgoing null cones ($l\propto\partial_r$), $n\simeq\partial_u-\tfrac12\partial_r$, and $m,\bar m$ tangent to the spheres. With $g^{ab}=-2l^{(a}n^{b)}+2m^{(a}\bar m^{b)}$, the contraction of a symmetric tensor $X_{cd}$ with the Hessian reads
\begin{align}
X_{cd}f^{cd}={}&X_{ll}f_{nn}+X_{nn}f_{ll}+2X_{ln}f_{ln}+2X_{m\bar m}f_{m\bar m}\notag\\
&-2\big(X_{l\bar m}f_{nm}+X_{n\bar m}f_{lm}+{\rm c.c.}\big)\notag\\
&+\big(X_{\bar m\bar m}f_{mm}+{\rm c.c.}\big).
\label{eq:contraction}
\end{align}
For a scalar with the radiative falloff $\Phi=\varphi_0+\varphi_1(u,x^A)/r+O(r^{-2})$ the tetrad components of the Hessian are
\begin{gather}
f_{nn}=\frac{f_1\ddot\varphi_1}{r}+O(r^{-2}),\qquad
f_{ln}\simeq f_{m\bar m}\simeq-\frac{f_1\dot\varphi_1}{r^2},\notag\\
f_{ll}=\frac{2f_1\varphi_1}{r^3}+O(r^{-4}),\qquad f_{nm}=O(r^{-2}),\notag\\
f_{lm},\,f_{mm}=O(r^{-3}),
\label{eq:hessian}
\end{gather}
where $f_1\equiv f'(\varphi_0)$ and a dot denotes $\partial_u$. The equality of $f_{ln}$ and $f_{m\bar m}$ at $O(r^{-2})$ is the cancellation that makes $\Box\Phi$ start at $O(r^{-3})$. The Weyl scalars peel, $\Psi_n=O(r^{n-5})$, and the Ricci part obeys $\hat S_{nn}=O(r^{-2})$, $\hat S_{ln},\hat S_{nm},\hat S_{m\bar m}=O(r^{-3})$, $\hat S_{ll},\hat S_{lm},\hat S_{mm}=O(r^{-4})$.

The key observation follows from Eq.~\eqref{eq:contraction}: the dominant Hessian component $f_{nn}$ multiplies only Weyl components with two $l$ indices, i.e.\ $\Psi_2$, $\Psi_1$ and $\Psi_0$. The radiative components $\Psi_4$ and $\Psi_3$ are annihilated by this contraction, because the radiation field is of type N (III) with repeated principal null direction $l$. The naive estimate $H\sim\Psi_4\,f_{nn}\sim r^{-2}$ is therefore never realized, and one finds
\begin{gather}
H_{nn},\ H_{nm},\ H_{mm}=O(r^{-4}),\notag\\
H_{ln},\ H_{m\bar m},\ H_{lm}=O(r^{-5}),\qquad H_{ll}=O(r^{-6}).
\label{eq:tetradorders}
\end{gather}
The products responsible for the leading terms are $f_{nn}\Psi_2$, $f_{nm}\Psi_3$, $f_{mm}\Psi_4$ and $\hat S_{ln}f_{nn}$, $\hat S_{nn}f_{ln}$ in $H_{nn}$; $f_{ln}\Psi_3$, $f_{m\bar m}\Psi_3$, $f_{lm}\Psi_4$ in $H_{nm}$; and the single product $f_{ll}\Psi_4$ in $H_{mm}$. The first products of tensor and scalar radiation thus appear two orders below the order relevant for the memory effect.

The trace-free angular component can be obtained in two lines and without any symmetry assumption. In Bondi coordinates $g^{uu}=0$ and $g^{ur}=-1+O(r^{-2})$, hence
\begin{gather}
f^{uu}=(g^{ur})^2f_{rr}=\frac{2f_1\varphi_1}{r^3}+O(r^{-4}),\notag\\
P_{AuBu}=-\frac r2\,\partial_u^2c_{AB}+O(r^0),
\end{gather}
and therefore
\begin{equation}
H^{\rm TF}_{AB}=-\frac{f_1\,\varphi_1\,\partial_uN_{AB}}{r^2}+O(r^{-3}),
\label{eq:HTF}
\end{equation}
where $c_{AB}$ is the shear and $N_{AB}=\partial_uc_{AB}$ the news. This is the product $f_{ll}\Psi_4$: the scalar charge aspect times the time derivative of the news. The intermediate steps, and the generalization to couplings with $f'(\varphi_0)=0$, are given in Appendix~\ref{app:derivations}.

\section{Asymptotic falloff of the Gauss--Bonnet stress tensor}
\label{sec:falloff}

We write the metric in Bondi--Sachs form~\cite{Bondi:1962px,Sachs:1962wk,Madler:2016xju},
\begin{align}
ds^2={}&-\frac Vr\,e^{2\beta}du^2-2e^{2\beta}du\,dr\notag\\
&+r^2h_{AB}\big(dx^A-U^Adu\big)\big(dx^B-U^Bdu\big),
\label{eq:BS}
\end{align}
with $\det h_{AB}=\det q_{AB}$, $q_{AB}$ the unit-sphere metric and $D_A$ its covariant derivative, and assume smooth expansions in $1/r$,
\begin{gather}
h_{AB}=q_{AB}+\frac{c_{AB}}{r}+\frac{d_{AB}}{r^2}+\frac{e_{AB}}{r^3}+\dots,\notag\\
\Phi=\varphi_0+\frac{\varphi_1}{r}+\frac{\varphi_2}{r^2}+\dots,\qquad
\beta=\frac{\beta_{(2)}}{r^2}+\dots,\notag\\
U^A=\frac{U^A_{(2)}}{r^2}+\frac{U^A_{(3)}}{r^3}+\dots,\quad V=r-2M+\frac{V_{(1)}}{r}+\dots
\label{eq:expansions}
\end{gather}
The Einstein--scalar equations fix, at orders where $\lambda$ does not appear (Sec.~\ref{sec:hierarchy}),
\begin{equation}
\beta_{(2)}=-\tfrac1{32}c_{AB}c^{AB}-\tfrac1{16}\varphi_1^2,\qquad U^A_{(2)}=-\tfrac12D_Bc^{AB},
\label{eq:leading}
\end{equation}
while $c_{AB}(u,x^A)$ and $\varphi_1(u,x^A)$ are free data and the mass aspect obeys
\begin{equation}
\partial_uM=-\tfrac18N_{AB}N^{AB}-\tfrac14\dot\varphi_1^{\,2}+\tfrac14D_AD_BN^{AB}.
\label{eq:massloss}
\end{equation}

Evaluating $H_{ab}$ on this background we find the falloffs collected in Table~\ref{tab:falloff}, where they are compared with the slowest falloff that the hierarchy tolerates for a generic effective source, Eq.~(2.15) of Ref.~\cite{Maibach:2026afb}.

\begin{table}[t]
\caption{Falloff of the coordinate components of $T^{\rm GB}_{ab}=-(\lambda/2\pi)H_{ab}$ in Bondi coordinates, compared with the bounds of Ref.~\cite{Maibach:2026afb} for a smooth $\mathscr{I}^+$. The last column is the number of powers of $1/r$ by which the Gauss--Bonnet term lies below the bound.}
\label{tab:falloff}
\begin{ruledtabular}
\begin{tabular}{lccc}
Component & $T^{\rm GB}$ & Bound & Margin\\
\hline
$T_{uu}$ & $r^{-4}$ & $r^{-2}$ & 2\\
$T_{ur}$ & $r^{-5}$ & $r^{-3}$ & 2\\
$T_{rr}$ & $r^{-6}$ & $r^{-3}$ & 3\\
$T_{uA}$ & $r^{-3}$ & $r^{-2}$ & 1\\
$T_{rA}$ & $r^{-4}$ & $r^{-2}$ & 2\\
$T_{AB}$ (trace-free part) & $r^{-2}$ & $r^{-1}$ & 1\\
$h^{AB}T_{AB}$ & $r^{-3}$ & $r^{-1}$ & 2\\
$g^{ab}T_{ab}$ & $r^{-5}$ & --- & ---\\
\end{tabular}
\end{ruledtabular}
\end{table}

Denoting by $H^{(n)}$ the coefficient of $r^{-n}$, the leading coefficients are
\begin{align}
H^{{\rm TF}(2)}_{AB}&=-f_1\,\varphi_1\,\partial_uN_{AB},\label{eq:HAB}\\
H^{(4)}_{rA}&=-f_1\,\varphi_1\,D^BN_{AB},\label{eq:HrA}\\
H^{(3)}_{uA}&=f_1\big[\partial_uN_{AB}\,D^B\varphi_1-\dot\varphi_1\,D^BN_{AB}\big],\label{eq:HuA}\\
H^{(5)}_{ur}&=f_1\Big[-\big(4M+\tfrac12c_{AB}N^{AB}\big)\dot\varphi_1\notag\\
&\qquad\ +D^A\varphi_1\,D^BN_{AB}\Big],\label{eq:Hur}\\
H^{(6)}_{rr}&=f_1\Big[-\big(4M+\tfrac12c_{AB}N^{AB}\big)\varphi_1\notag\\
&\qquad\ +\tfrac12\varphi_1^2\dot\varphi_1\Big],\label{eq:Hrr}
\end{align}
and
\begin{align}
H^{(4)}_{uu}={}&f_1\Big[-2\,\partial_u\big(M\dot\varphi_1\big)-\tfrac14\partial_u\big(c_{AB}N^{AB}\dot\varphi_1\big)\notag\\
&-\tfrac18\varphi_1\partial_u\big(N_{AB}N^{AB}\big)-\tfrac12\dot N^{AB}D_AD_B\varphi_1\notag\\
&+\tfrac12\dot\varphi_1D_AD_BN^{AB}+D_A\dot\varphi_1\,D_BN^{AB}\Big].\label{eq:Huu}
\end{align}
Using Eq.~\eqref{eq:massloss}, the last expression can be rewritten without $\partial_uM$, which generates a term $\tfrac12f_1\dot\varphi_1^{\,3}$. The traces are
\begin{align}
\big(h^{AB}H_{AB}\big)^{(3)}={}&f_1\Big[-\big(8M+c_{AB}N^{AB}\big)\dot\varphi_1\notag\\
&+2D^A\varphi_1D^BN_{AB}-\tfrac12\varphi_1^2\ddot\varphi_1\notag\\
&-2\varphi_1\dot\varphi_1^{\,2}\Big],\label{eq:trang}\\
\big(g^{ab}H_{ab}\big)^{(5)}={}&-\tfrac12f_1\varphi_1\big(\varphi_1\ddot\varphi_1+4\dot\varphi_1^{\,2}\big).\label{eq:tr}
\end{align}

Several remarks are in order. First, the coefficients depend only on the leading data $(c_{AB},M,\varphi_1)$ and on $f_1$; no subleading coefficient of the metric or of the scalar appears, and neither does $f''(\varphi_0)$. Second, the combination $4M+\tfrac12c_{AB}N^{AB}$ equals $-4\Psi_2^0$ up to a term quadratic in the scalar, cf.\ Eq.~\eqref{eq:psis}, in agreement with the counting of Sec.~\ref{sec:structure}. Third, all terms containing $N_{AB}$ are genuine cross terms between the tensor and scalar radiative sectors. Finally, Eq.~\eqref{eq:tr} agrees with an independent evaluation of the second identity in Eq.~\eqref{eq:identities}.

Equations~\eqref{eq:HAB}--\eqref{eq:tr} hold for a general Bondi--Sachs metric with both polarizations of the shear and arbitrary dependence on $(u,\theta,\phi)$, and involve no subleading coefficient of the metric or of the scalar. They were obtained covariantly and confirmed by an independent order-by-order integration of the full field equations, as described in Appendix~\ref{app:checks}.

\section{Bondi--Sachs hierarchy}
\label{sec:hierarchy}

\subsection{Orders at which the coupling enters}

The Bondi--Sachs field equations split into hypersurface equations for $\beta$, $U^A$ and $V$, an evolution equation for the trace-free part of $h_{AB}$, and supplementary equations that yield the balance laws for the mass aspect and the angular-momentum aspect~\cite{Bondi:1962px,Sachs:1962wk,Madler:2016xju,Flanagan:2015pxa,Maibach:2026afb}. A source enters them through $T_{rr}$, $T_{rA}$, the trace combination, $T^{\rm TF}_{AB}$, and $T_{uu}$, $T_{uA}$, respectively. Combining this structure with Table~\ref{tab:falloff} one obtains Table~\ref{tab:hierarchy}, which we have confirmed by computing the full equation \eqref{eq:tracerev} order by order.

\begin{table*}[t]
\caption{First order in $1/r$ at which $\lambda$ appears in each field equation, and the coefficient determined at that order. The orders at which the leading data are fixed are listed for comparison. The supplementary equations at $O(r^{-2})$, which give the balance laws for the mass and angular-momentum aspects, are unaffected.}
\label{tab:hierarchy}
\begin{ruledtabular}
\begin{tabular}{llcl}
Equation & Orders without $\lambda$ & First order with $\lambda$ & Affects\\
\hline
$R_{rr}$ (hypersurface, $\beta$) & $r^{-3}$--$r^{-5}$: $\beta_{(1)},\beta_{(2)},\beta_{(3)}$ & $r^{-6}$ & $\beta_{(4)}$\\
$R_{rA}$ (hypersurface, $U^A$) & $r^{-1}$--$r^{-3}$: $U_{(1)},U_{(2)}$, log condition & $r^{-4}$ & $U^A_{(4)}$\\
$R_{ur}$ (hypersurface, $V$) & $r^{-4}$: $V_{(1)}$ & $r^{-5}$ & $V_{(2)}$\\
$R^{\rm TF}_{AB}$ (evolution) & $r^{0}$, $r^{-1}$ & $r^{-2}$ & $\partial_ue^{\rm TF}_{AB}$\\
$R_{uu}$ (supplementary) & $r^{-2}$, $r^{-3}$ & $r^{-4}$ & ---\\
$R_{uA}$ (supplementary) & $r^{-2}$ & $r^{-3}$ & ---\\
\end{tabular}
\end{ruledtabular}
\end{table*}

For the hypersurface functions the $\lambda$-dependent contributions follow directly from the coefficients of Sec.~\ref{sec:falloff}:
\begin{gather}
\beta_{(4)}\supset\tfrac14\lambda\,H^{(6)}_{rr}=\lambda f_1\varphi_1\Big[-\big(M+\tfrac18c_{AB}N^{AB}\big)+\tfrac18\varphi_1\dot\varphi_1\Big],\notag\\
U^A_{(4)}\supset-2\lambda\,H^{(4)}_{rB}\,q^{AB}=2\lambda f_1\,\varphi_1\,D_BN^{AB}.
\label{eq:betaU}
\end{gather}

\subsection{Smoothness to all orders}
\label{sec:smooth}

In a $1/r$ expansion a logarithm is forced whenever the coefficient multiplying the highest unknown of an equation vanishes, so that the equation turns into a constraint on previously determined data. The coefficients of the unknowns are fixed by the radial operators of the hierarchy, which are those of GR. At the order where each unknown is determined one finds (Appendix~\ref{app:derivations})
\begin{gather}
R_{rr}\big|_{r^{-(n+2)}}\supset-4n\,\beta_{(n)},\notag\\
R_{rA}\big|_{r^{-n}}\supset\tfrac12n(n-3)\,U_{A(n)},\notag\\
R_{ur}\big|_{r^{-(n+3)}}\supset\tfrac12n(n+1)\,V_{(n)},\notag\\
R^{\rm TF}_{AB}\big|_{r^{1-k}}\supset-(k-1)\,\partial_uh^{(k)}_{AB},\notag\\
\Box\Phi\big|_{r^{-(k+1)}}\supset2(k-1)\,\partial_u\varphi_k,
\label{eq:radialops}
\end{gather}
where $h^{(k)}_{AB}$ is the coefficient of $r^{-k}$ in $h_{AB}$ and $U_{A(n)}=q_{AB}U^B_{(n)}$. The resonant orders are therefore $\beta_{(0)}$, $V_{(0)}=-2M$, $h^{(1)}_{AB}=c_{AB}$, $\varphi_1$ and $U^A_{(3)}$. The first four are the gauge condition and the free data; the last is the integration function associated with the angular-momentum aspect, and the equation $R_{rA}$ at $O(r^{-3})$ becomes the integrability condition that removes logarithms, namely the vanishing of the trace-free part of $d_{AB}$~\cite{Chrusciel:1993hx,Winicour:1985}.

The Gauss--Bonnet source never reaches these orders. The resonant equations involve $T_{rr}$ at $O(r^{-2})$, $T_{rA}$ at $O(r^{-3})$, $T_{ur}$ and $g^{ab}T_{ab}$ at $O(r^{-3})$ (equivalently $h^{AB}T_{AB}$ at $O(r^{-1})$), $T^{\rm TF}_{AB}$ at $O(r^{0})$ and the scalar source at $O(r^{-2})$, while the Gauss--Bonnet contributions start at $O(r^{-6})$, $O(r^{-4})$, $O(r^{-5})$ (respectively $O(r^{-3})$), $O(r^{-2})$ and $O(r^{-6})$ (Table~\ref{tab:falloff} and Sec.~\ref{sec:scalar}). At every other order the unknown has a nonvanishing coefficient and is determined algebraically, or through a $u$-integration, by a source that is a polynomial in previously determined coefficients and their derivatives; $H_{ab}$ and $\mathcal{G}$ are polynomials in the metric and scalar coefficients and in the expansions of $e^{\pm2\beta}$ and $h^{AB}$, so no logarithm can be generated. By induction, if the free data are smooth and the trace-free part of $d_{AB}$ vanishes, the expansion is smooth to all orders in $1/r$, for arbitrary $f(\Phi)$.

In particular, the Gauss--Bonnet term enters $R_{rA}$ only at $O(r^{-4})$ and the trace-free angular equation only at $O(r^{-2})$, one order beyond the conditions (2.25) and (2.44) of Ref.~\cite{Maibach:2026afb}. This settles the status of sGB gravity with respect to the classification of that reference. The logarithms of the Horndeski map of the Gauss--Bonnet term~\cite{Kobayashi:2011nu} do not translate into logarithms of the metric: they cancel among the Horndeski functionals, as is manifest in the covariant form \eqref{eq:metric}. In the axisymmetric integration below, the condition reads $\gamma_{2,\theta}+2\cot\theta\,\gamma_2=0$, and the scalar contributions to it, $8\beta_{2,\theta}$ and $\varphi_1\varphi_{1,\theta}$, cancel identically.

\subsection{Explicit integration in the axisymmetric gauge}

To integrate the hierarchy explicitly we specialize to the axisymmetric, twist-free metric of Bondi, van der Burg and Metzner~\cite{Bondi:1962px},
\begin{align}
ds^2={}&-\Big(\frac Vre^{2\beta}-U^2r^2e^{2\gamma}\Big)du^2-2e^{2\beta}du\,dr\notag\\
&-2Ur^2e^{2\gamma}du\,d\theta\notag\\
&+r^2\big(e^{2\gamma}d\theta^2+e^{-2\gamma}\sin^2\!\theta\,d\phi^2\big),
\end{align}
with $\gamma=c/r+\gamma_2/r^2+\gamma_3/r^3+\dots$, so that $c_{\theta\theta}=2c$, $c_{\phi\phi}=-2c\sin^2\!\theta$, $c_{AB}N^{AB}=8c\dot c$, $D^Bc_{B\theta}=2(c_\theta+2c\cot\theta)$ and $D^AD^Bc_{AB}=2(c_{\theta\theta}+3c_\theta\cot\theta-2c)$, where subscripts $\theta$ denote derivatives. Keeping all coefficients of $\gamma$, $\beta$, $U$, $V$ and $\Phi$ as free functions of $(u,\theta)$ and solving Eq.~\eqref{eq:tracerev} order by order we find
\begin{gather}
\beta_1=0,\qquad\beta_2=-\tfrac14c^2-\tfrac1{16}\varphi_1^2,\qquad\beta_3=-\tfrac16\varphi_1\varphi_2,\notag\\
U_1=0,\qquad U_2=-(c_\theta+2c\cot\theta),\notag\\
\gamma_{2,\theta}+2\cot\theta\,\gamma_2=0\ \Rightarrow\ \gamma_2=0,
\label{eq:lowsol}
\end{gather}
the last line being the condition that removes logarithms from the expansion ($U_3$ decouples at that order and remains free). Following Bondi we parametrize the subleading data as
\begin{equation}
U_3=2N+3c\,c_\theta+4c^2\cot\theta,\qquad\gamma_3=C-\tfrac16c^3 .
\end{equation}

\subsubsection{Balance laws}

The supplementary equation $R_{uu}$ at $O(r^{-2})$ gives
\begin{equation}
\partial_uM=-\dot c^{\,2}-\tfrac14\dot\varphi_1^{\,2}+\tfrac12\partial_u\big(c_{\theta\theta}+3c_\theta\cot\theta-2c\big),
\end{equation}
which is Eq.~\eqref{eq:massloss}, and $R_{u\theta}$ at $O(r^{-2})$ gives
\begin{align}
-3\,\partial_uN={}&\partial_\theta M+3c\,\dot c_\theta+4c\,\dot c\cot\theta+\dot c\,c_\theta\notag\\
&+\tfrac18\varphi_1\dot\varphi_{1,\theta}-\tfrac38\varphi_{1,\theta}\dot\varphi_1 .
\label{eq:Nevol}
\end{align}
Neither contains $\lambda$. For $\varphi_1=0$ they reduce to the classical equations of Ref.~\cite{Bondi:1962px}, and the scalar term in the mass loss has the coefficient found in Refs.~\cite{Lu:2019jus,Tan:2020hfa}. Because the field equations are linear in $\lambda$ and the coefficients \eqref{eq:HAB}--\eqref{eq:Huu} depend only on data fixed at $\lambda$-independent orders, these statements are exact in $\lambda$; the only assumption is the smooth expansion \eqref{eq:expansions}. In covariant terms they follow from $T^{\rm GB}_{uu}=O(r^{-4})$ and $T^{\rm GB}_{uA}=O(r^{-3})$, which lie below the orders $r^{-2}$ at which a source contributes to the balance laws~\cite{Flanagan:2015pxa,Maibach:2026afb}. Consequently, the flux-balance laws associated with supertranslations and with the Lorentz generators, and the displacement and spin memory formulas derived from them~\cite{Strominger:2014pwa,Pasterski:2015tva}\cite{Nichols:2017rqr}, coincide with those of the Einstein--scalar system. This reproduces by direct integration the conclusions of Refs.~\cite{Heisenberg:2023prj,Maibach:2026jbf} and extends them to the angular-momentum sector.

\subsubsection{Evolution of the quadrupole-order datum}

The trace-free angular equation at $O(r^{-2})$ determines $\partial_u\gamma_3$. In Bondi's variables,
\begin{align}
4\,\partial_uC={}&2c^2\dot c+2cM+N\cot\theta-\partial_\theta N\notag\\
&+\tfrac18\big(\varphi_1\varphi_{1,\theta\theta}-\cot\theta\,\varphi_1\varphi_{1,\theta}-\varphi_{1,\theta}^2\big)
-8\lambda f_1\,\varphi_1\,\ddot c .
\label{eq:Cevol}
\end{align}
The first four terms are the vacuum equation of Ref.~\cite{Bondi:1962px}, the bracket is the contribution of the minimally coupled scalar, and the last term is the Gauss--Bonnet contribution. Since the linear part of the Einstein operator is the same with or without symmetry and the source is known covariantly, Eq.~\eqref{eq:HAB}, the new term reads in general
\begin{equation}
\partial_ue^{\rm TF}_{AB}\Big|_\lambda=-2\lambda\,f'(\varphi_0)\,\varphi_1\,\partial_uN_{AB},
\label{eq:eevol}
\end{equation}
where $e^{\rm TF}_{AB}$ is the trace-free part of the $O(r^{-3})$ coefficient of $h_{AB}$. This agrees with the general evolution equation (2.46) of Ref.~\cite{Maibach:2026afb} upon inserting $T^{(2)}_{AB}=(\lambda/2\pi)f_1\varphi_1\partial_uN_{AB}$ from Eq.~\eqref{eq:HAB}, and with the radial operator of Eq.~\eqref{eq:radialops} at $k=3$.

Equation \eqref{eq:Cevol} is the first dynamical equation of the hierarchy in which the coupling appears. The function $C$ is the quadrupole-order datum of Bondi's analysis and corresponds, in the Newman--Penrose language~\cite{Newman:1961qr,Newman:1962cia}, to the leading coefficient of $\Psi_0$ (Sec.~\ref{sec:NP}). The asymptotic data are organized as $M\leftrightarrow\Psi_2$, $N_A\leftrightarrow\Psi_1$, $e_{AB}\leftrightarrow\Psi_0$: the Gauss--Bonnet coupling is invisible in the first two levels and appears in the third. The new term vanishes if the scalar has no $1/r$ part or if the news is constant in time. Integrated between two nonradiative epochs, its net effect is
\begin{equation}
\Delta e^{\rm TF}_{AB}\Big|_\lambda=-2\lambda f_1\int du\,\varphi_1\,\partial_uN_{AB}=2\lambda f_1\int du\,\dot\varphi_1\,N_{AB},
\label{eq:Deltae}
\end{equation}
since the boundary term vanishes when the news switches off: a correlation between the scalar and tensor news. This is the counterpart, in the presence of scalar radiation, of the observation of Ref.~\cite{Lu:2020mjp} that in the $D\to4$ limit the Gauss--Bonnet term acts one order after the integration functions and in the quadrupole.

At the following orders we find
\begin{align}
\beta_4={}&-\tfrac34c\,\gamma_3-\tfrac3{16}\varphi_1\varphi_3-\tfrac18\varphi_2^2\notag\\
&+\lambda f_1\varphi_1\Big[-(M+c\dot c)+\tfrac18\varphi_1\dot\varphi_1\Big],\\
U_4\supset{}&4\lambda f_1\varphi_1\big(\dot c_\theta+2\dot c\cot\theta\big),\\
V_2\supset{}&\lambda f_1\Big[8(M+c\dot c)\dot\varphi_1-4\varphi_{1,\theta}\big(\dot c_\theta+2\dot c\cot\theta\big)\Big],
\label{eq:V2}
\end{align}
in agreement with Eq.~\eqref{eq:betaU}; the expression for $V_2$ includes the contributions induced by $\beta_4$ and $U_4$, and $\supset$ denotes the complete $\lambda$-dependent part.

\subsection{Newman--Penrose form}
\label{sec:NP}

To express the result in terms of Weyl scalars we use the null tetrad adapted to the Bondi frame,
\begin{gather}
l=e^{-2\beta}\partial_r,\qquad n=\partial_u-\frac{V}{2r}\partial_r+U^A\partial_A,\notag\\
m=\frac{E_1+iE_2}{\sqrt2\,r},
\label{eq:tetrad}
\end{gather}
where $E_1=h_{\theta\theta}^{-1/2}\partial_\theta$ and $E_2=h_{\theta\theta}^{1/2}(\sin\theta)^{-1}\big(\partial_\phi-h_{\theta\phi}h_{\theta\theta}^{-1}\partial_\theta\big)$ form an orthonormal frame of $h_{AB}$; in the axisymmetric gauge $m=(\sqrt2\,r)^{-1}\big(e^{-\gamma}\partial_\theta+i\,e^{\gamma}\csc\theta\,\partial_\phi\big)$. With $l\cdot n=-1$ and $m\cdot\bar m=1$ we define
\begin{gather}
\Psi_0=C_{abcd}l^am^bl^cm^d,\quad \Psi_1=C_{abcd}l^an^bl^cm^d,\notag\\
\Psi_2=C_{abcd}l^am^b\bar m^cn^d,\quad
\Psi_3=C_{abcd}l^an^b\bar m^cn^d,\notag\\
\Psi_4=C_{abcd}n^a\bar m^bn^c\bar m^d .
\end{gather}
Because $g_{rr}=g_{rA}=0$ and $g_{AB}m^Am^B=0$, one has exactly $\Psi_0=(l^r)^2R_{rArB}\,m^Am^B$: $\Psi_0$ is the trace-free part of $R_{rArB}$. At linear order a term $h^{(n)}_{AB}r^{-n}$ of $h_{AB}$ contributes $-\tfrac12n(n-1)h^{(n)}_{AB}r^{-n}$ to it, so that the shear drops out and
\begin{equation}
\Psi_0^0=-3\,e_{AB}\,\hat m^A\hat m^B+(\text{nonlinear terms}),
\label{eq:Psi0lin}
\end{equation}
with $\hat m^A\equiv r\,m^A$, where the nonlinear terms involve only the leading data. Evaluating the Weyl scalars on the asymptotic solution of the axisymmetric gauge we find the peeling behavior $\Psi_n=\Psi_n^0\,r^{n-5}+O(r^{n-6})$, with
\begin{gather}
\Psi_4^0=-\ddot c,\qquad
\Psi_3^0=-\tfrac{1}{\sqrt2}\big(\dot c_\theta+2\dot c\cot\theta\big),\notag\\
\Psi_2^0=-M-c\,\dot c-\tfrac1{12}\varphi_1\dot\varphi_1,\notag\\
\Psi_1^0=\tfrac{3}{\sqrt2}N-\tfrac{\sqrt2}{16}\varphi_1\varphi_{1,\theta},\qquad
\Psi_0^0=-6\,C-\tfrac14c\,\varphi_1^2 .
\label{eq:psis}
\end{gather}
Covariantly, $\Psi_4^0=-\tfrac12\partial_uN_{AB}\,\hat{\bar m}^A\hat{\bar m}^B$ and $\Psi_3^0=-\tfrac12D^BN_{AB}\,\hat{\bar m}^A$. None of these coefficients depends on $\lambda$.

Taking the retarded-time derivative of $\Psi_0^0$ and using the field equations through $O(r^{-2})$ in the trace-free angular equation, we obtain, without any symmetry assumption,
\begin{equation}
\begin{aligned}
\partial_u\Psi_0^0={}&\eth\Psi_1^0+3\sigma^0\Psi_2^0-\tfrac14\varphi_1\eth^2\varphi_1+\tfrac12(\eth\varphi_1)^2\\
&-\tfrac14\varphi_1\partial_u\big(\sigma^0\varphi_1\big)-12\lambda f'(\varphi_0)\,\varphi_1\,\bar\Psi_4^0 ,
\end{aligned}
\label{eq:Psi0gen}
\end{equation}
where $\sigma^0=\tfrac12c_{AB}\hat m^A\hat m^B$ is the asymptotic shear, so that $\Psi_4^0=-\partial_u^2\bar\sigma^0$, and $\eth$ acts on a quantity of spin weight $s$ as $\eth\eta=\tfrac1{\sqrt2}\big(\partial_\theta+i\csc\theta\,\partial_\phi-s\cot\theta\big)\eta$. In the axisymmetric gauge $\sigma^0=c$ and Eq.~\eqref{eq:Psi0gen} follows directly from Eqs.~\eqref{eq:massloss}, \eqref{eq:Nevol} and \eqref{eq:Cevol}. The first two terms are the vacuum asymptotic Bianchi identity of Newman and Unti~\cite{Newman:1962cia}, which our computation reproduces identically; the next three are the contribution of the minimally coupled scalar (analogous matter terms appear in the Einstein--Maxwell--dilaton analysis of Ref.~\cite{Tan:2020hfa}); the last term is the Gauss--Bonnet contribution. It has the spin weight $+2$ of $\Psi_0$, as it must, and it is precisely the product $f_{ll}\Psi_4$ identified in Sec.~\ref{sec:structure}.

The Gauss--Bonnet term can also be read off from Eqs.~\eqref{eq:eevol} and \eqref{eq:Psi0lin}: the nonlinear terms in $\Psi_0^0$ involve only the free data $c_{AB}$ and $\varphi_1$, so that $\lambda$ enters $\partial_u\Psi_0^0$ only through $\partial_ue_{AB}$,
\begin{align}
\partial_u\Psi_0^0\Big|_\lambda&=6\lambda f'(\varphi_0)\,\varphi_1\,\partial_uN_{AB}\,\hat m^A\hat m^B\notag\\
&=-12\lambda f'(\varphi_0)\,\varphi_1\,\bar\Psi_4^0 .
\label{eq:Psi0lam}
\end{align}
By contrast, the evolution equations of $\Psi_2^0$ and $\Psi_1^0$, equivalent to Eqs.~\eqref{eq:massloss} and \eqref{eq:Nevol}, are those of the Einstein--scalar system. In the language of asymptotic charges, the mass (spin-0) and angular-momentum (spin-1) aspects are blind to the Gauss--Bonnet coupling, and the first aspect whose evolution is modified is the spin-2 one built from $\Psi_0$~\cite{Freidel:2021ytz}.

\subsection{Scalar hierarchy}
\label{sec:scalar}

On the asymptotic solution the Gauss--Bonnet invariant decays as $r^{-6}$. In the axisymmetric gauge, and using the mass-loss formula, $\mathcal{G}=\mathcal{G}_6/r^6+O(r^{-7})$ with
\begin{align}
\mathcal{G}_6={}&48\,(M+c\dot c)^2+96\,C\,\ddot c+96\,N\big(\dot c_\theta+2\dot c\cot\theta\big)\notag\\
&+8\,(M+c\dot c)\,\varphi_1\dot\varphi_1+4\,c\,\ddot c\,\varphi_1^2\notag\\
&-4\,\varphi_1\varphi_{1,\theta}\big(\dot c_\theta+2\dot c\cot\theta\big)-\varphi_1^2\dot\varphi_1^{\,2}.
\label{eq:G6}
\end{align}
The first three terms display the structure $\Psi_2^2$, $\Psi_0\Psi_4$, $\Psi_1\Psi_3$ of the square of the Weyl tensor; for Schwarzschild one recovers $\mathcal{G}=48M^2/r^6$. Expanding Eq.~\eqref{eq:scalar}, the equation at $O(r^{-(k+1)})$ contains $2(k-1)\partial_u\varphi_k$:
\begin{align}
O(r^{-3}):&\ \ 2\,\partial_u\varphi_2+D^2\varphi_1=0,\notag\\
O(r^{-4}):&\ \ 4\,\partial_u\varphi_3+(D^2+2)\varphi_2-2M\varphi_1+\dots=0,\notag\\
O(r^{-5}):&\ \ 6\,\partial_u\varphi_4+(D^2+6)\varphi_3-8M\varphi_2+\dots=0,\notag\\
O(r^{-6}):&\ \ 8\,\partial_u\varphi_5+(D^2+12)\varphi_4-18M\varphi_3+\dots\notag\\
&\ \ =-\lambda f_1\mathcal{G}_6,
\label{eq:scalarhier}
\end{align}
where the dots denote products of the shear and its derivatives with lower scalar coefficients, and the $O(r^{-2})$ equation is identically satisfied, so that $\varphi_1(u,x^A)$ is free data. The coupling therefore enters the scalar hierarchy at the evolution of $\varphi_5$, both through $\mathcal{G}_6$ and through $V_2$ and $U_4$. The scalar news $\dot\varphi_1$ remains free, and the value of $\varphi_1$ in a stationary epoch (the scalar charge) is not fixed by the asymptotic analysis.

As a check, in the static limit ($c=N=C=0$, no $u$ dependence) Eqs.~\eqref{eq:scalarhier} give $\varphi_2=M\varphi_1$, $\varphi_3=\tfrac43M\varphi_2$ and $12\varphi_4-18M\varphi_3=-48\lambda f_1M^2$ up to terms of higher order in the scalar. With $\varphi_4=0$, as selected by regularity at the horizon, one recovers the perturbative hair on a Schwarzschild background~\cite{Sotiriou:2014pfa,Yagi:2011xp} \cite{Yunes:2011we},
\begin{equation}
\Phi-\varphi_0=\frac{2\lambda f_1}{Mr}\Big(1+\frac Mr+\frac{4M^2}{3r^2}\Big)+O(\lambda^2)\textcolor{blue}{,}
\label{eq:hair}
\end{equation}
which solves Eq.~\eqref{eq:scalar} on the Schwarzschild background exactly at first order in $\lambda$.

\section{Spin-2 balance law, BMS properties and the tower of couplings}
\label{sec:charges}

\subsection{Balance law of the spin-2 aspect}
\label{sec:spin2}

Equation~\eqref{eq:Psi0gen} has the structure of the spin-2 member of the recursion obeyed by the tower of higher-spin charge aspects of Freidel, Pranzetti and Raclariu~\cite{Freidel:2021ytz}, in which the evolution of the spin-$s$ aspect is sourced by $\eth$ of the spin-$(s-1)$ aspect and by the shear times the spin-$(s-2)$ aspect; here it is supplemented by the scalar and Gauss--Bonnet fluxes. Integrating it between two nonradiative epochs $u_1<u_2$ gives the balance law
\begin{equation}
\Delta\Psi_0^0=\int_{u_1}^{u_2}\!du\,\Big[\eth\Psi_1^0+3\sigma^0\Psi_2^0+\mathcal{S}_\varphi\Big]+\mathcal{F}_{\rm GB},
\label{eq:balance}
\end{equation}
where $\mathcal{S}_\varphi$ collects the three scalar terms of Eq.~\eqref{eq:Psi0gen} and
\begin{equation}
\mathcal{F}_{\rm GB}=-12\lambda f_1\!\int\! du\,\varphi_1\bar\Psi_4^0=-6\lambda f_1\!\int\! du\,\dot\varphi_1\,N_{AB}\hat m^A\hat m^B .
\label{eq:FGB}
\end{equation}
The second form follows from $\bar\Psi_4^0=-\tfrac12\partial_uN_{AB}\hat m^A\hat m^B$ after an integration by parts, with no boundary term because the news vanishes in nonradiative epochs. It is the spin-2 projection of Eq.~\eqref{eq:Deltae}, $\mathcal{F}_{\rm GB}=-3\,\Delta e_{AB}\hat m^A\hat m^B|_\lambda$.

Two properties make $\mathcal{F}_{\rm GB}$ a meaningful observable even though $\Psi_0^0$ is not. Under a supertranslation generated by $T(x^A)$ the radiative data transform homogeneously, $\delta_TN_{AB}=T\,\partial_uN_{AB}$ and $\delta_T\varphi_1=T\dot\varphi_1$, at the order relevant here, and the leading asymptotic structure on which these transformations rest is that of the Einstein--scalar system, as established in Sec.~\ref{sec:hierarchy}. Hence
\begin{equation}
\delta_T\mathcal{F}_{\rm GB}=-6\lambda f_1\int du\,\partial_u\big(T\,\dot\varphi_1N_{AB}\hat m^A\hat m^B\big)=0 .
\label{eq:Tinv}
\end{equation}
By contrast, the transformation of $\Psi_0^0$ contains $\eth T\,\Psi_1^0$ and higher powers of $\eth T$ multiplying $\Psi_2^0$, $\Psi_3^0$ and $\Psi_4^0$, so that $\Delta\Psi_0^0$, and with it $\Delta e_{AB}$, depends on the BMS frame; in the gauge-invariant reading of Eq.~\eqref{eq:balance} all the frame dependence resides in the Einstein--scalar terms. Under Lorentz transformations both $\varphi_1$ and $N_{AB}$ transform with definite conformal weights and no inhomogeneous term, and the terms proportional to $u\,\partial_u$ integrate by parts, so that $\mathcal{F}_{\rm GB}$ transforms covariantly. Under superrotations the news acquires a $u$-independent shift proportional to $D_{\langle A}D_{B\rangle}D_CY^C$~\cite{Barnich:2010eb,Compere:2018ylh}, and $\mathcal{F}_{\rm GB}$ shifts by a term proportional to the net change $\Delta\varphi_1$ of the scalar charge aspect, mirroring the vacuum-transition behavior of the displacement memory. The Gauss--Bonnet flux is thus a correlation between the scalar and tensor news with the same frame properties as the nonlinear memory flux, and it is the leading imprint of the coupling on the asymptotic charges.

The quantities entering Eq.~\eqref{eq:balance} are accessible to numerical relativity: Cauchy-characteristic extraction delivers the full set of Weyl scalars at $\mathscr{I}^+$~\cite{Moxon:2020gha}, and balance laws of this type are routinely used to compute memory and to test waveforms~\cite{Mitman:2020pbt,Mitman:2020bjf}. In sGB simulations~\cite{Witek:2018dmd,Gasparotto:2026,Hu:2026qnm,Corman:2025sgb}, Eq.~\eqref{eq:balance} provides a nonperturbative consistency condition that involves the coupling explicitly, through a flux built entirely from radiative data.

\subsection{Couplings with vanishing derivatives and scalarized branches}
\label{sec:tower}

The counting of Sec.~\ref{sec:structure} extends to couplings whose first derivatives vanish at the asymptotic value of the scalar. Suppose that $f^{(k)}(\varphi_0)=0$ for $1\le k<n$ and $f_n\equiv f^{(n)}(\varphi_0)\neq0$. Since $\Phi-\varphi_0=\varphi_1/r+O(r^{-2})$ uniformly in all directions, every tetrad component of the Hessian is shifted by $n-1$ powers of $1/r$, the dominant trace-free angular contribution is again $f_{ll}\Psi_4$, and (Appendix~\ref{app:derivations})
\begin{equation}
H^{\rm TF}_{AB}=-\frac{(n+1)\,f_n\,\varphi_1^n\,\partial_uN_{AB}}{2\,(n-1)!\;r^{n+1}}+O(r^{-n-2}).
\label{eq:towerH}
\end{equation}
For $n=2$, with $f^{uu}=3f''(\varphi_0)\varphi_1^2/r^4$, this is
\begin{equation}
H^{\rm TF}_{AB}\Big|_{f_1=0}=-\frac{3f''(\varphi_0)\,\varphi_1^2\,\partial_uN_{AB}}{2r^3}+O(r^{-4}).
\label{eq:Hn2}
\end{equation}
All lower orders of the hierarchy are those of the Einstein--scalar system, and the radial operator of Eq.~\eqref{eq:radialops} at $k=n+2$ gives
\begin{equation}
\partial_uh^{(n+2){\rm TF}}_{AB}\Big|_\lambda=-\frac{2\lambda f_n}{(n-1)!}\,\varphi_1^n\,\partial_uN_{AB}.
\label{eq:towerh}
\end{equation}
Writing $\Psi_0=\sum_{j\ge0}\Psi_0^j\,r^{-5-j}$, the linear relation $\Psi_0^{k-3}\supset-\tfrac12k(k-1)h^{(k)}_{AB}\hat m^A\hat m^B$ of Sec.~\ref{sec:NP} yields
\begin{equation}
\partial_u\Psi_0^{n-1}\Big|_\lambda=-\frac{2(n+1)(n+2)}{(n-1)!}\,\lambda f_n\,\varphi_1^n\,\bar\Psi_4^0,
\label{eq:towerPsi}
\end{equation}
while $\Psi_0^j$ with $j<n-1$ evolves as in the Einstein--scalar system. For $n=1$ this is Eq.~\eqref{eq:Psi0lam}; for $n=2$ the source is $-24\lambda f''(\varphi_0)\varphi_1^2\bar\Psi_4^0$ in the evolution of $\Psi_0^1$, the coefficient of $r^{-6}$, which is the level at which the Newman--Penrose quantities~\cite{Newman:1968uj} and the subleading BMS charges~\cite{Godazgar:2018dvh} reside, and which carries the spin-3 aspect of the higher-spin tower~\cite{Freidel:2021ytz}. In general, the order of the first nonvanishing derivative of the coupling function selects the spin, $s=n+1$, of the first charge aspect that feels the Gauss--Bonnet term. The case $n=1$ includes the dilatonic and shift-symmetric couplings; $n=2$ includes the quadratic and exponential-quadratic couplings of spontaneous scalarization~\cite{Doneva:2017bvd,Silva:2017uqg,Doneva:2022ewd}.

The size of the effect depends on which branch of solutions the spacetime belongs to. For $n=1$ and black holes in vacuum, the scalar is sourced by the coupling itself, Eq.~\eqref{eq:hair} gives $\varphi_1=O(\lambda)$, and all Gauss--Bonnet contributions to the hierarchy are $O(\lambda^2)$. This is the asymptotic counterpart of the well-known fact that metric corrections on the branch connected to GR start at second order in the coupling~\cite{Yunes:2011we,Yagi:2011xp}, and it implies that there is no $O(\lambda)$ correction to the memory formulas or to the leading asymptotic data on that branch. For $n=2$ the situation is different. The GR solutions, with $\varphi_1=0$, remain solutions and carry no Gauss--Bonnet imprint at all, while on the scalarized branches the scalar charge is fixed by the nonlinear structure of the solution and is not suppressed by an additional power of the coupling~\cite{Doneva:2022ewd}, and in binaries it can grow dynamically once the separation drops below a critical value~\cite{Capuano:2026hair}. There Eq.~\eqref{eq:towerPsi} is a first-order effect in $\lambda$ controlled by a nonperturbative charge, entering one step higher in the tower than for $n=1$. The new terms are also of first order in $\lambda$ whenever the $1/r$ part of the scalar does not vanish with $\lambda$, as for matter sources carrying their own scalar charge or for independent scalar radiation.

\subsection{Leading-order estimate for a black-hole binary}
\label{sec:binary}

To illustrate the size and pattern of $\mathcal{F}_{\rm GB}$ consider a quasicircular binary of nonspinning black holes with $n=1$, masses $m_{1,2}$, total mass $M$, reduced mass $\mu$, separation $a$ and orbital frequency $\Omega$, at leading post-Newtonian order. Each hole carries the scalar charge $D_a=2\lambda f_1/m_a$ of Eq.~\eqref{eq:hair}, which we take constant along the inspiral at this order; the adiabatic evolution of the charges, and their dynamical growth in scalarizing models~\cite{Capuano:2026hair}, enter at higher order. The scalar and tensor radiative data follow from the multipole expansions~\cite{Thorne:1980ru,Blanchet:2013haa}, $\varphi_1=\sum_\ell(\ell!)^{-1}n_L\,\partial_u^\ell\mathcal{D}^L$ with $\mathcal{D}^L=\sum_aD_ax_a^L$, and $c_{AB}=2\ddot I_{ij}e^i_Ae^j_B$ with $I_{ij}$ the mass quadrupole. For equal masses the scalar dipole and all odd multipoles vanish, and the only non-oscillating contribution to $\dot\varphi_1N_{AB}$ at leading order is the product of the scalar quadrupole $\mathcal{D}^{ij}=\mu_{\mathcal{D}}a^2\hat n^i\hat n^j$, with $\mu_{\mathcal{D}}=(D_1m_2^2+D_2m_1^2)/M^2$, and the mass quadrupole $I^{ij}=\mu a^2\hat n^i\hat n^j$, both radiating at $2\Omega$ and in phase. Averaging over an orbit,
\begin{equation}
\big\langle\dot\varphi_1N_{AB}\hat m^A\hat m^B\big\rangle=4\,\mu_{\mathcal{D}}\,\mu\,a^4\Omega^6\sin^2\iota\,\big(1+\cos^2\iota\big),
\end{equation}
where $\iota$ is the angle between the orbital axis and the direction of observation. The flux therefore grows secularly, like a memory. Integrating along the GR quadrupole inspiral, $da/du=-\tfrac{64}{5}\mu M^2/a^3$, from large separation down to $a_f$ and using $m_1=m_2$, for which $\mu_{\mathcal{D}}=2\lambda f_1/M$, Eq.~\eqref{eq:FGB} gives
\begin{equation}
\mathcal{F}_{\rm GB}=-\frac{15}{4}\,\frac{\lambda^2f_1^2}{a_f}\,\sin^2\iota\,\big(1+\cos^2\iota\big).
\label{eq:binary}
\end{equation}
The effect is quadratic in the coupling, as required by the $n=1$ counting, it is dominated by the late inspiral, and its angular pattern is that of the nonlinear displacement memory. For unequal masses the scalar dipole combines with the mass octupole and current quadrupole at the orbital frequency and contributes at the same post-Newtonian order, and the merger adds a contribution that requires the full nonlinear evolution~\cite{Witek:2018dmd,Shiralilou:2020gah,Shiralilou:2021mfl,Julie:2019sab}.

\section{Conclusions}
\label{sec:conclusions}

We have determined where the coupling of scalar--Gauss--Bonnet gravity enters the Bondi--Sachs description of radiating spacetimes, and through which asymptotic charge it first becomes visible. The structural reason for its late appearance is that the dominant component of the Hessian of the scalar is contracted, in the double-dual Riemann tensor, with the repeated principal null direction of the radiative curvature, which annihilates the radiative Weyl scalars. The effective stress tensor then lies between one and three inverse powers of the radial coordinate below the falloffs that would affect the leading asymptotic structure, with explicit covariant coefficients that hold without any symmetry assumption.

Three consequences follow. The balance laws for the mass and angular-momentum aspects, and with them the displacement and spin memories, are those of the Einstein--scalar system exactly in the coupling. The smooth expansion at null infinity persists to all orders for an arbitrary coupling function, because the Gauss--Bonnet source never reaches the resonant orders of the radial integrations; scalar--Gauss--Bonnet gravity, although its Horndeski representation contains logarithmic functionals, belongs to the class of theories with a smooth null infinity. And the coupling leaves its first imprint in the balance law of the spin-2 aspect, as a flux that correlates the scalar charge aspect with the time derivative of the news. This flux is invariant under supertranslations and covariant under Lorentz transformations, so that it isolates the Gauss--Bonnet contribution from the frame-dependent part of the spin-2 charge.

These results complement the existing literature in three ways. They provide the explicit Bondi--Sachs solution that underlies the statements of Refs.~\cite{Heisenberg:2023prj,Maibach:2026jbf} on the absence of Gauss--Bonnet contributions to the memory and to the supertranslation balance law, and extend them to the angular-momentum sector. They show that scalar--Gauss--Bonnet gravity, although outside the class of regular functionals considered in Ref.~\cite{Maibach:2026afb}, does not belong to the logarithmic case: treated covariantly, it satisfies all the falloff bounds with margin. And they identify the level of the asymptotic hierarchy, that of the spin-2 aspect, at which a higher-curvature coupling of this type leaves its first imprint, generalizing the $D\to4$ result of Ref.~\cite{Lu:2020mjp} to a theory with scalar radiation.

The same mechanism organizes the whole family of coupling functions into a tower: the order of the first nonvanishing derivative of the coupling at the asymptotic value of the scalar fixes the spin of the first charge aspect that responds to it, with a closed-form source. This makes the distinction between branches of solutions sharp. On the branch connected to general relativity the imprint is quadratic in the coupling, while for the couplings responsible for spontaneous scalarization it is linear in the coupling and controlled by nonperturbative scalar charges, one step higher in the tower. For black-hole binaries the flux accumulates secularly during the inspiral with the angular pattern of the nonlinear memory, which makes it a natural target for waveforms extracted at null infinity.

The analysis opens several directions. The construction of Wald--Zoupas charges and fluxes at the spin-2 and spin-3 levels would place the balance law derived here within a symplectic framework, and its evaluation on numerical sGB waveforms extracted at null infinity would turn it into a quantitative test of the coupling. The same counting applies to the axion--Pontryagin coupling of dynamical Chern--Simons gravity~\cite{Hou:2021oxe}, for which an equally late appearance in the hierarchy is expected.

\appendix

\section{Intermediate algebra}
\label{app:derivations}

We collect here the steps behind Eqs.~\eqref{eq:HTF}, \eqref{eq:radialops} and \eqref{eq:towerH}--\eqref{eq:towerPsi}. Let $f(\Phi)-f(\varphi_0)=\sum_{k\ge n}f_k(\Phi-\varphi_0)^k/k!$ with $f_n\neq0$. In Bondi gauge $\Gamma^u_{rr}=\Gamma^A_{rr}=0$ and $\Gamma^r_{rr}=2\partial_r\beta=O(r^{-3})$, so that $f_{rr}=\partial_r^2f+O(r^{-n-4})$, and from $(\Phi-\varphi_0)^n=\varphi_1^n r^{-n}+O(r^{-n-1})$ one finds
\begin{equation}
f_{rr}=\frac{(n+1)\,f_n\,\varphi_1^n}{(n-1)!\;r^{n+2}}+O(r^{-n-3}).
\end{equation}
Since $g^{uu}=g^{uA}=0$ exactly, $f^{uu}=(g^{ur})^2f_{rr}=e^{-4\beta}f_{rr}$, and $l=e^{-2\beta}\partial_r$ gives $f_{ll}=e^{-4\beta}f_{rr}$; for $n=1$ this reproduces the entry $2f_1\varphi_1/r^3$ of Eq.~\eqref{eq:hessian}. All other tetrad components of the Hessian are obtained from Eq.~\eqref{eq:hessian} in the same way and are shifted uniformly by $n-1$ powers of $1/r$, so that the counting of Eq.~\eqref{eq:tetradorders} is shifted by the same amount and the dominant trace-free angular term is again $f_{ll}\Psi_4$.

The Weyl component that multiplies $f^{uu}$ in $H_{AB}$ is $P_{AuBu}$. Its Ricci part, $-g_{AB}R_{uu}+g_{Au}R_{uB}+g_{uB}R_{Au}-g_{uu}R_{AB}+\tfrac12R(g_{AB}g_{uu}-g_{Au}g_{Bu})$, is either proportional to $g_{AB}$ or of order $r^{-2}$ in its trace-free part, because on shell $R^{\rm TF}_{AB}=\tfrac12(\partial_A\varphi_1\partial_B\varphi_1)^{\rm TF}r^{-2}+O(r^{-3})$. The trace-free part of $P_{AuBu}$ is therefore that of $R_{AuBu}$. On the background \eqref{eq:BS}, the only term of $R_{AuBu}$ of order $r$ comes from $-\tfrac12\partial_u^2g_{AB}$, which gives $-\tfrac r2\partial_u^2c_{AB}=-\tfrac r2\partial_uN_{AB}$. Multiplying by $f^{uu}$ yields Eqs.~\eqref{eq:HTF} and \eqref{eq:towerH}.

The coefficient of $\partial_uh^{(k)}_{AB}$ in $R_{AB}$ follows from the terms of the Ricci tensor that contain $\partial_ug_{AB}$ on the background $g_{ur}=-1$, $g_{AB}=r^2q_{AB}$. With $g_{AB}\supset r^{2-k}h^{(k)}_{AB}$ one has $\Gamma^u_{AB}=\tfrac12\partial_rg_{AB}$, $\Gamma^r_{AB}\supset\tfrac12\partial_ug_{AB}$, $\Gamma^C_{uB}=\tfrac12g^{CD}\partial_ug_{DB}$ and $\Gamma^C_{rA}=r^{-1}\delta^C_A+O(r^{-2})$. The contributions of $\partial_u\Gamma^u_{AB}$, $\partial_r\Gamma^r_{AB}$, $\Gamma^c_{cr}\Gamma^r_{AB}$, $-\Gamma^C_{Ar}\Gamma^r_{CB}-\Gamma^r_{AC}\Gamma^C_{rB}$ and $-\Gamma^u_{AC}\Gamma^C_{uB}-\Gamma^C_{Au}\Gamma^u_{CB}$ are, in units of $r^{1-k}\partial_uh^{(k)}_{AB}$, equal to $\tfrac12(2-k)$, $\tfrac12(2-k)$, $1$, $-1$ and $-1$, which add up to the coefficient $-(k-1)$ quoted in Eq.~\eqref{eq:radialops}. Combined with $R^{\rm TF}_{AB}=-4\lambda H^{\rm TF}_{AB}+\dots$ from Eq.~\eqref{eq:tracerev}, this gives $\partial_uh^{(k){\rm TF}}_{AB}|_\lambda=4\lambda H^{{\rm TF}(k-1)}_{AB}/(k-1)$, which reproduces Eq.~\eqref{eq:eevol} for $k=3$ and Eq.~\eqref{eq:towerh} for $k=n+2$.

The remaining radial operators in Eq.~\eqref{eq:radialops} are obtained in the same way from $R_{rr}=4r^{-1}\partial_r\beta+\dots$, from the term $\tfrac12r^{-2}\partial_r\big(r^4\partial_rU_A\big)$ of $R_{rA}$, from the term $\tfrac12r^{-1}\partial_r^2V$ of $R_{ur}$ and from the term $-2r^{-1}\partial_u\partial_r(r\Phi)$ of $\Box\Phi$, evaluated on monomials $r^{-n}$; the vanishing of $\tfrac12n(n-3)$ at $n=3$ is the origin of the integrability condition discussed in Sec.~\ref{sec:smooth}. Finally, the coefficient $-\tfrac12k(k-1)$ relating $h^{(k)}_{AB}$ to $\Psi_0$ follows from $R_{rArB}=-\tfrac12\partial_r^2g_{AB}+\tfrac14g^{CD}\partial_rg_{AC}\partial_rg_{BD}+\dots$, whose linearization on $g_{AB}=r^2(q_{AB}+h^{(k)}_{AB}r^{-k})$ gives $\big[(2-k)-1-\tfrac12(2-k)(1-k)\big]h^{(k)}_{AB}r^{-k}=-\tfrac12k(k-1)h^{(k)}_{AB}r^{-k}$.

\section{Explicit integration of the hierarchy}
\label{app:checks}

The expressions of Secs.~\ref{sec:falloff} and \ref{sec:hierarchy} were derived with SymPy~\cite{Meurer:2017yhf}. The metric functions and the scalar are written as Laurent series in $1/r$ with arbitrary coefficient functions, and the Christoffel symbols, the Riemann tensor, the double dual \eqref{eq:P}, the Hessian of $f(\Phi)$ and the contraction \eqref{eq:metric} are then expanded component by component, keeping in each case the orders needed for the hierarchy. Two identities served as a guide throughout. The relations \eqref{eq:identities} hold exactly on a static, spherically symmetric metric with arbitrary functions, which fixes the normalization of the Gauss--Bonnet term in Eq.~\eqref{eq:metric}. On the radiative background, the divergence $\nabla^aH_{ab}$ reproduces $-\tfrac18f_1\dot\varphi_1\mathcal{G}_6$ at $O(r^{-7})$ for $b=u$, which ties together the leading and next-to-leading coefficients of all components of $H_{ab}$.

We also integrated the complete field equations \eqref{eq:scalar} and \eqref{eq:tracerev} directly, without using any of the expressions of the main text, so that every result could be confirmed by a calculation of different nature. In this second approach the coefficient functions are represented by their local Taylor expansions in $(u,\theta,\phi)$, the free data $c_{AB}$ and $\varphi_1$ and the initial values of $M$, $U^A_{(3)}$, $\gamma_3$ and $\varphi_k$ are left generic, and the hierarchy is solved order by order: the hypersurface equations fix $\beta$, $U^A$ and $V$ algebraically, while the evolution equations determine the dependence on $u$ of $M$, $U^A_{(3)}$, the coefficients of $h_{AB}$ and the $\varphi_k$. The radial operators of Eq.~\eqref{eq:radialops} appear in this way as the coefficients of the unknowns at each order.

In the axisymmetric gauge the direct integration reproduces Eqs.~\eqref{eq:lowsol}, the mass-loss formula, Eqs.~\eqref{eq:Nevol} and \eqref{eq:Cevol}, the expressions for $\beta_4$, $U_4$ and $V_2$, the Weyl scalars \eqref{eq:psis} with their peeling, the evolution equation \eqref{eq:Psi0gen}, the invariant $\mathcal{G}_6$, the scalar hierarchy \eqref{eq:scalarhier} and the coefficients \eqref{eq:HAB}--\eqref{eq:tr}; comparing with the solution at $\lambda=0$ for the same free data isolates the Gauss--Bonnet contributions. Without symmetry assumptions, for both polarizations of the shear, it confirms Eqs.~\eqref{eq:HAB}--\eqref{eq:tr}, the covariant form of Eq.~\eqref{eq:Psi0gen}, the relation $\Psi_4^0=-\partial_u^2\bar\sigma^0$ and Eq.~\eqref{eq:eevol}. For couplings with $f'(\varphi_0)=0$ it confirms Eqs.~\eqref{eq:towerH} and \eqref{eq:towerh} for $n=2$ and $n=3$, and Eq.~\eqref{eq:towerPsi} for $n=2$, together with the fact that the lower coefficients evolve as in the Einstein--scalar system. Finally, Eq.~\eqref{eq:hair} solves Eq.~\eqref{eq:scalar} on the Schwarzschild background at every order, and the orbital average behind Eq.~\eqref{eq:binary} follows directly from the quadrupole waveforms.

\bibliography{references}

\end{document}